\documentclass[twocolumn]{aastex631}
\usepackage{hyperref}
\usepackage{amsmath}
\usepackage{amssymb}

\newcommand{\fargo}{\texttt{FARGO3D}}
\newcommand{\idefix}{\texttt{Idefix}}
\newcommand{\athena}{\texttt{Athena++}}
\newcommand{\pluto}{\texttt{PLUTO}}
\newcommand{\Phantom}{\texttt{Phantom}}
\newcommand{\mcfost}{\texttt{mcfost}}
\newcommand{\radmc}{\texttt{RADMC-3D}}
\newcommand{\discminer}{\texttt{DISCMINER}}

\shorttitle{exoALMA benchmark}
\shortauthors{Bae et al.}

\begin{document}

\title{exoALMA VII: Benchmarking Hydrodynamics and Radiative Transfer Codes}

\correspondingauthor{Jaehan Bae}
\email{jbae@ufl.edu}

\author[0000-0001-7258-770X]{Jaehan Bae}
\affiliation{Department of Astronomy, University of Florida, Gainesville, FL 32611, USA}

\author[0000-0002-9298-3029]{Mario Flock}
\affiliation{Max-Planck Institute for Astronomy (MPIA), Königstuhl 17, 69117 Heidelberg, Germany}

\author[0000-0001-8446-3026]{Andres Izquierdo}
\altaffiliation{NASA Hubble Fellowship Program Sagan Fellow}
\affiliation{Leiden Observatory, Leiden University, P.O. Box 9513, NL-2300 RA Leiden, The Netherlands}
\affiliation{European Southern Observatory, Karl-Schwarzschild-Str. 2, D-85748 Garching bei M\"nchen, Germany}
\affiliation{Department of Astronomy, University of Florida, Gainesville, FL 32611, USA}

\author[0000-0001-7235-2417]{Kazuhiro Kanagawa}
\affiliation{College of Science, Ibaraki University, 2-1-1 Bunkyo, Mito, Ibaraki 310-8512, Japan}

\author[0000-0001-8524-6939]{Tomohiro Ono}
\affiliation{School of Natural Sciences, Institute for Advanced Study, Princeton, NJ 08544, USA}

\author[0000-0001-5907-5179]{Christophe Pinte}
\affiliation{Univ. Grenoble Alpes, CNRS, IPAG, 38000 Grenoble, France}
\affiliation{School of Physics and Astronomy, Monash University, VIC 3800, Australia}

\author[0000-0002-4716-4235]{Daniel J. Price}
\affiliation{School of Physics and Astronomy, Monash University, VIC 3800, Australia}

\author[0000-0003-4853-5736]{Giovanni P. Rosotti}
\affiliation{Dipartimento di Fisica, Universit\`a degli Studi di Milano, Via Celoria 16, I-20133 Milano, Italy}

\author[0000-0002-3468-9577]{Gaylor Wafflard-Fernandez}
\affiliation{Univ. Grenoble Alpes, CNRS, IPAG, 38000 Grenoble, France}

\author[0000-0002-8896-9435]{Geoffroy Lesur}
\affiliation{Univ. Grenoble Alpes, CNRS, IPAG, 38000 Grenoble, France}

\author[0000-0002-9626-2210]{Frederic Masset}
\affiliation{Instituto de Ciencias Físicas, Universidad Nacional Autonoma de México, Av. Universidad s/n, 62210 Cuernavaca, Mor., México}

\author[0000-0003-2253-2270]{Sean M. Andrews}
\affiliation{Center for Astrophysics | Harvard \& Smithsonian, Cambridge, MA 02138, USA}

\author[0000-0001-6378-7873]{Marcelo Barraza-Alfaro}
\affiliation{Department of Earth, Atmospheric, and Planetary Sciences, Massachusetts Institute of Technology, Cambridge, MA 02139, USA}

\author[0000-0000-0000-0000]{Myriam Benisty}
\affiliation{Univ. Grenoble Alpes, CNRS, IPAG, 38000 Grenoble, France}
\affiliation{Universit\'e C\^ote d'Azur, Observatoire de la C\^ote d'Azur, CNRS, Laboratoire Lagrange, France}

\author[0000-0002-2700-9676]{Gianni Cataldi}
\affiliation{National Astronomical Observatory of Japan, 2-21-1 Osawa, Mitaka, Tokyo 181-8588, Japan}

\author[0000-0003-3713-8073]{Nicol\'as Cuello}
\affiliation{Univ. Grenoble Alpes, CNRS, IPAG, 38000 Grenoble, France}

\author[0000-0003-2045-2154]{Pietro Curone}
\affiliation{Dipartimento di Fisica, Università degli Studi di Milano, via Celoria 16, 20133 Milano, Italy}
\affiliation{Departamento de Astronomía, Universidad de Chile, Camino El Observatorio 1515, Las Condes, Santiago, Chile}

\author[0000-0002-1483-8811]{Ian Czekala}
\affiliation{School of Physics \& Astronomy, University of St. Andrews, North Haugh, St. Andrews KY16 9SS, UK}

\author[0000-0003-4689-2684]{Stefano Facchini}
\affiliation{Dipartimento di Fisica, Università degli Studi di Milano, via Celoria 16, 20133 Milano, Italy}

\author[0000-0003-4679-4072]{Daniele Fasano}
\affiliation{Université Côte d'Azur, Observatoire de la Côte d'Azur, CNRS, Laboratoire Lagrange, France}

\author[0000-0002-5503-5476]{Maria Galloway-Sprietsma}
\affiliation{Department of Astronomy, University of Florida, Gainesville, FL 32611, USA}

\author[0000-0002-8138-0425]{Cassandra Hall}
\affiliation{Department of Physics and Astronomy, The University of Georgia, Athens, GA 30602, USA}
\affiliation{Center for Simulational Physics, The University of Georgia, Athens, GA 30602, USA}
\affiliation{Institute for Artificial Intelligence, The University of Georgia, Athens, GA, 30602, USA}

\author[0000-0003-1502-4315]{Iain Hammond} 
\affiliation{School of Physics and Astronomy, Monash University, VIC 3800, Australia}

\author[0000-0001-6947-6072]{Jane Huang}
\affiliation{Department of Astronomy, Columbia University, 538 W. 120th Street, Pupin Hall, New York, NY 10027, USA}

\author[0000-0002-2357-7692]{Giuseppe Lodato} 
\affiliation{Dipartimento di Fisica, Universit\`a degli Studi di Milano, Via Celoria 16, 20133 Milano, Italy}

\author[0000-0003-4663-0318]{Cristiano Longarini} 
\affiliation{Institute of Astronomy, University of Cambridge, Madingley Road, CB3 0HA, Cambridge, UK}
\affiliation{Dipartimento di Fisica, Universit\`a degli Studi di Milano, Via Celoria 16, 20133 Milano, Italy}

\author[0000-0002-0491-143X]{Jochen Stadler}
\affiliation{Universit\'{e} C\^{o}te d'Azur, Observatoire de la C\^{o}te d'Azur, CNRS, Laboratoire Lagrange, 06300 Nice, France}
\affiliation{Univ. Grenoble Alpes, CNRS, IPAG, 38000 Grenoble, France}

\author[0000-0003-1534-5186]{Richard Teague}
\affiliation{Department of Earth, Atmospheric, and Planetary Sciences, Massachusetts Institute of Technology, Cambridge, MA 02139, USA}

\author[0000-0003-1526-7587]{David Wilner}
\affiliation{Center for Astrophysics | Harvard \& Smithsonian, Cambridge, MA 02138, USA}

\author[0000-0002-7501-9801]{Andrew Winter}
\affiliation{Universit\'{e} C\^{o}te d'Azur, Observatoire de la C\^{o}te d'Azur, CNRS, Laboratoire Lagrange, 06300 Nice, France}

\author[0000-0002-7212-2416]{Lisa W\"olfer} 
\affiliation{Department of Earth, Atmospheric, and Planetary Sciences, Massachusetts Institute of Technology, Cambridge, MA 02139, USA}

\author[0000-0001-8002-8473]{Tomohiro C. Yoshida}
\affiliation{National Astronomical Observatory of Japan, 2-21-1 Osawa, Mitaka, Tokyo 181-8588, Japan}
\affiliation{Department of Astronomical Science, The Graduate University for Advanced Studies, SOKENDAI, 2-21-1 Osawa, Mitaka, Tokyo 181-8588, Japan}



\begin{abstract}

Forward modeling is often used to interpret substructures observed in protoplanetary disks. To ensure the robustness and consistency of the current forward modeling approach from the community, we conducted a systematic comparison of various hydrodynamics and radiative transfer codes. Using four grid-based hydrodynamics codes (\fargo, \idefix, \athena, \pluto) and a smoothed particle hydrodynamics code (\Phantom), we simulated a protoplanetary disk with an embedded giant planet. We then used two radiative transfer codes (\mcfost, \radmc) to calculate disk temperatures and create synthetic $^{12}$CO cubes. Finally, we retrieved the location of the planet from the synthetic cubes using \discminer. We found strong consistency between the hydrodynamics codes, particularly in the density and velocity perturbations associated with planet-driven spirals. We also found a good agreement between the two radiative transfer codes: the disk temperature in \mcfost\ and \radmc\ models agrees within $\lesssim 3~\%$ everywhere in the domain. In synthetic $^{12}$CO channel maps, this results in brightness temperature differences within $\pm1.5$~K in all our models. This good agreement ensures consistent retrieval of planet's radial/azimuthal location with only a few \% of scatter, with velocity perturbations varying $\lesssim 20~\%$ among the models. Notably, while the planet-opened gap is shallower in the \Phantom\ simulation, we found that this does not impact the planet location retrieval. In summary, our results demonstrate that any combination of the tested hydrodynamics and radiative transfer codes can be used to reliably model and interpret planet-driven kinematic perturbations.

\end{abstract}




\section{Introduction} 
\label{sec:intro}

Recent high-resolution observations have revealed a range of substructures in protoplanetary disks, including gaps, rings, crescents, and spirals in dust continuum and velocity kinks in molecular line channel maps. While unequivocally connecting these substructures to their cause(s) is not straightforward \citep[see e.g., reviews by][]{andrews2020,bae2023}, one exciting possibility is that embedded protoplanets have created the observed substructures. 

As protoplanets grow, they perturb their natal protoplanetary disks, producing observable signatures. While these perturbations manifest in various forms, in this paper we focus on kinematic perturbations observable with radio interferometric molecular line observations. In the vicinity of a protoplanet, spiral arms launched by the protoplanet induce strong perturbations in the density, temperature, and velocity of the background disk. These perturbations can manifest as velocity kinks in channel maps and/or Doppler flips in velocity centroid maps (see review by \citealt{pinte2023} and references therein). Because these perturbations are generally strongest near protoplanets, velocity kinks and Doppler flips offer an opportunity to locate the protoplanet. In a more global scale, spirals launched by a protoplanet exchange angular momentum with the disk gas, causing an annular gap to open around the protoplanet's orbit \citep{lin1993}. The radially varying pressure gradient associated with the gap modulates the azimuthal velocity of the disk gas. As a result, inward to the protoplanet's orbit, the gas rotates at a sub-Keplerian speed while outward to the orbit, the gas rotate's at a super-Keplerian speed \citep{kanagawa2015}. Because the width and depth of a planet-opened gap depend on the mass of the embedded protoplanet, the observed azimuthal velocity of the disk gas can be used to infer the protoplanet's mass \citep{zhang2018,yun2019}.  

Hydrodynamic planet-disk interaction simulations, combined with radiative transfer calculations, can give us deeper understanding of the nature of observed substructures. For example, this forward modeling approach enables one to infer the location and mass of the planet responsible for observed substructures \citep[e.g.,][]{pinte2018,bae2019,bae2021}, which can offer crucial insights into planet formation processes.  However, while there had been previous efforts to compare properties of the gap a planet opens and the gravitational torque a planet exerts on its protoplanetary disks using various hydrodynamics codes \citep{devalborro2006}, to date, no systematic hydrodynamics or radiative transfer benchmark tests comparing kinematic perturbations induced by planets embedded in their birth protoplanetary disks have been carried out. To this end, in this paper we tested if various hydrodynamics and radiative transfer codes produce consistent results. We benchmarked five hydrodynamics codes, namely \fargo\ \citep{fargo3d}, \idefix\ \citep{idefix}, \athena\ \citep{athena}, \pluto\ \citep{pluto}, and \Phantom\ \citep{Phantom}, and two radiative transfer codes, \mcfost\ \citep{Pinte2006,Pinte2009} and \radmc\ \citep{radmc}, which are some of the most widely used codes from the community. Our benchmark test considered a protoplanetary disk with an embedded giant planet having a mass that is $10^{-3}$ of the central star, which corresponds to about a Jupiter mass in a system with a solar-mass central star.

This paper is organized as follows. In Section \ref{sec:hydro}, we describe the planet-disk interaction simulation setup and compare the hydrodynamics simulation results. In Section \ref{sec:RT}, we describe the radiative transfer simulation setup and compare disk temperatures and synthetic $^{12}$CO cubes. We retrieve the planet location in the synthetic cubes using \discminer\ and present the results in Section \ref{sec:discminer}. We summarize our findings and conclude in Section \ref{sec:conclusion}.

\section{Hydrodynamic Simulations} 
\label{sec:hydro}

\subsection{Grid-based Hydrodynamics Simulation Setup}

Grid-based simulations were carried out using four hydrodynamics codes: \fargo\ version 2.01\footnote{\url{https://github.com/FARGO3D/fargo3d}} \citep{fargo3d}, \idefix\
 version 1.1.0\footnote{\url{https://github.com/idefix-code/idefix/releases/tag/v1.1}} \citep{idefix}, \athena\ version 21.0\footnote{\url{https://github.com/PrincetonUniversity/athena/releases/tag/v21.0-dev}} \citep{athena}, and \pluto\ version 4.3\footnote{\url{https://plutocode.ph.unito.it}} \citep{pluto}. 
While we refer the reader to the relevant papers listed above for technical details of these codes, we emphasize their fundamental differences in solving hydrodynamic equations. Among the grid-based codes, \fargo\ employs a finite-difference scheme, which directly approximates derivatives in the differential form of conservation equations. The other three codes, namely \idefix, \athena, and \pluto, utilize a finite-volume scheme, which works with the integral form of conservation equations and computes inter-cell fluxes through Riemann solvers. These methodological differences can influence how each code handles discontinuities between grid cells and conserves physical quantities. It is thus crucial to verify that results obtained with these codes agree with each other.

For all the grid-based simulations presented in this paper, we used the orbital advection algorithm FARGO \citep[Fast Advection in Rotating Gaseous Objects;][]{fargo}, which enables us to have a much larger time step by subtracting the azimuthally-averaged velocity, although we ran additional simulations without implementing FARGO and confirmed that the outcome of the simulations do not depend on the use of the FARGO algorithm.

\subsubsection{Simulation Domain \& Numerical Resolution}
\label{sec:simulation_domain}

We used three-dimensional spherical coordinates $(r, \theta, \phi)$. Throughout this paper, we use $r$, $\theta$, and $\phi$ to denote the spherical radius, meridional angle, and azimuthal angle in spherical coordinates, while we use $R \equiv r \sin \theta$ and $Z \equiv r \cos \theta$ to denote the cylindrical radius and height in cylindrical coordinates. The simulation domain extends from  $r_{\rm in} = 0.3$ to $r_{\rm out} = 3.0$ in $r$ in the code unit, from $\theta_{\rm min} = \pi/2 - \arctan(0.5)$ to $\theta_{\rm max} = \pi/2 + \arctan(0.5)$ in $\theta$, and from $\phi_{\rm min} = -\pi$ to $\phi_{\rm max} = \pi$ in $\phi$. We used 288 logarithmically-spaced grid cells in $r$, 128 linear grid cells in $\theta$, and 784 linear grid cells in $\phi$, respectively. With this setup, we achieved $\Delta r \simeq r \Delta\theta \simeq r \Delta \phi  \simeq  0.008$ at the radial location of the planet ($r=1$), a numerical resolution allowing about 12 grid cells per scale height, which was shown to be sufficient for numerical convergence \citep{chen2024}.

\subsubsection{Initial Disk Setup}
\label{sec:initial_setup}

The disk temperature has a power-law profile as a function of the cylindrical radius $R$, with a slope $q=-0.5$:
\begin{equation}
\label{eqn:temp}
T(R) = T_p \left( R / R_p \right)^q.
\end{equation}
Here, $R_p$ is the radial distance between the star and the planet and $T_p$ is the temperature at $R=R_p$. We chose $T_p$ such that the disk aspect ratio $H/R$ at $R=R_p$ is 0.1, where $H$ denotes the disk scale height. This disk aspect ratio is close to the aspect ratio at the radial location of the planetary candidates identified in \citet{Pinte_exoALMA} when we adopt the temperature profile inferred in \citet{Galloway_exoALMA}. We assumed that the disk is vertically isothermal and adopted an isothermal equation of state. Under this assumption, the gas pressure $P$, gas density $\rho$, and sound speed $c_s$ satisfy $P = \rho c_s^2$. 

In reality, temperature structure of protoplanetary disks is more complex than the simplified temperature structure we adopted. The disk temperature is vertically stratified as the stellar irradiation diminishes toward the disk midplane \citep{chiang1997,dalessio1998}. In addition, spirals driven by a forming planet can produce shock/compressional heating \citep{richert2015,lyra2016} and/or can cast shadows, modifying the stellar radiation field \citep{muley2024}, all of which can alter the underlying temperature structure. However, we expect the difference in the velocity perturbations at the vicinity of the planet in isothermal simulations and those in more realistic simulations to be rather minor as shown in \citet{pinte2019} where planet-driven velocity perturbations are compared between an isothermal simulation and one where the disk temperature is regularly updated by running radiative transfer calculations.

We initialized the gas density to satisfy the vertical force balance:
\begin{equation}
\label{eqn:density}
    \rho(R,Z) = \rho_p \left( {R \over R_p} \right)^p \exp\left({{GM_* \over c_s^2}\left[ {1 \over \sqrt{R^2+Z^2}}- {1 \over R}\right]}\right).
\end{equation}
Here, $G$ is the gravitational constant and $M_*$ is the stellar mass.
We chose $\rho_p$ such that the total initial disk mass is $0.01~M_\odot$. Note however that the hydrodynamic simulations we carried out are scalable and the outcome does not depend on the choice of the disk mass; the disk mass only becomes relevant in radiative transfer calculations. We adopted $p=-2.25$ such that the surface density follows $\Sigma \propto R^{-1}$.

The angular velocity $\Omega$ was initialized to satisfy the radial force balance:
\begin{equation}
    \Omega (R,Z) = \Omega_K \left[ (p+q) \left( {H \over R} \right)^2  + (1+q) - {qR \over \sqrt{R^2 + Z^2}} \right]^{1/2},
\end{equation}
where 
$\Omega_K = \sqrt{GM_*/R^3}$ is the Keplerian angular velocity.
The initial azimuthal velocity was set to $v_\phi = R \Omega$. The initial radial and meridional velocities were set to zero ($v_r = v_\theta = 0$).

We adopted kinematic viscosity characterized by $\nu = 10^{-5}$. In terms of the Shakura–Sunyaev viscosity parameter $\alpha$ \citep{shakura1973}, this kinematic viscosity is equivalent to $\alpha=10^{-3}$ at the radial location of the planet.

\subsubsection{Planet Potential}
\label{sec:planet_potential}

We considered a planet having a mass $M_p = 10^{-3}M_*$. For a solar-mass star, this planet mass corresponds to about a Jupiter mass, which is close to the lower limit we expect to detect in exoALMA-quality data \citep{Pinte_exoALMA}. The planet potential $\Phi_p$ is smoothed following
\begin{equation}
    \Phi_p = -{G M_p \over d} \left[ \left( {d \over \epsilon} \right)^4 -2 \left( {d \over \epsilon} \right)^3 +2 \left( {d \over \epsilon} \right) \right]
\end{equation}
when $d \leq \epsilon$, where $d$ is the distance between the planet and the grid cell in question, and $\epsilon = 3\Delta r \simeq 0.024$, where $\Delta r \simeq 0.008$ is the size of the grid cell at the location of the planet. When $d > \epsilon$, $\Phi_p = - GM_p/d$.

The mass of the planet was ramped up following 
\begin{equation}
    M_p(t) = M_{p,{\rm final}} \sin^2 \left( {t \over \tau_{\rm ramp}} {\pi \over 2} \right),
\end{equation}
where $M_{p, {\rm final}}=10^{-3}~M_*$ and $\tau_{\rm ramp} = 100$ orbits at the radial location of the planet. No gas accretion onto the planet is allowed, so the planet mass is fixed to $10^{-3}~M_*$ after 100 orbits.
We fixed the planet on a circular orbit and did not allow orbital migration.

\subsubsection{Boundary Conditions}

At the inner and outer radial boundaries, we adopted the symmetric boundary condition, although the choice of radial boundary condition should not have a significant impact on the outcome because we implemented a wave-killing zone at the radial boundaries
\citep{devalborro2006}. The damping function was written as 
\begin{equation}
    {dX(r) \over dt} = -\left( {{X(r)-X_0(r)} \over \tau_{\rm damp}} \right) f(r),
\end{equation}
where $X$ is density or velocity and $X_0$ is the initial value of the variable $X$. The inner damping zone was located between $r_1 = r_{\rm in} = 0.3$ and $r_2 = 0.357$, where we set the function $f(r)$ at the inner radial boundary $f_{\rm in}(r)$ to
\begin{equation}
    f_{\rm in}(r) = 1-\sin^2\left[ {\pi \over 2} {(r-r_1) \over (r_2-r_1)}\right].
\end{equation} 
The outer damping zone was located between $r_1 = 2.52$ and $r_2 = r_{\rm out} = 3.0$. The function $f(r)$ at the outer radial boundary, $f_{\rm out}(r)$, is given by
\begin{equation}
    f_{\rm out}(r) = \sin^2\left[ {\pi \over 2} {(r-r_1) \over (r_2-r_1)}\right].
\end{equation}
The damping timescale $\tau_{\rm damp}$ at the inner/outer radial boundary was given by $1\%$ of the orbital timescale at the inner/outer edge of the domain.

At the meridional boundaries $\theta = \theta_{\rm min}$ and $\theta = \theta_{\rm max}$, we adopted the symmetric boundary condition for $v_r$ and $v_\phi$.  We adopted the anti-symmetric boundary condition for $v_\theta$ such that $v_\theta = 0$ at the meridional boundary and that the meridional velocity in the active zone and the meridional velocity in the ghost zone, referring to a few extra cells beyond the boundary of the simulation domain, satisfy $v_\theta({\rm ghost}) = -v_\theta({\rm active})$. The density in the ghost zone was set to satisfy the vertical hydrostatic equilibrium \citep{bae2016}:
\begin{equation}
\label{eqn:boundary}
    {1 \over \rho} {\partial \over \partial \theta} (\rho c_s^2) = {v_\phi^2 \over \tan \theta}.
\end{equation}

\subsection{Smoothed Particle Hydrodynamics Simulation Setup}

The smoothed particle hydrodynamics (SPH) simulation was carried out using the code \Phantom\ version 2022.0.1\footnote{\url{https://github.com/danieljprice/phantom/releases/tag/v2022.0.1}} \citep{Phantom}. \Phantom\ solves the equations of hydrodynamics in Lagrangian form. The fluid is discretized onto a set of particles of mass and thus \Phantom\ is mesh-free by construction. We refer the reader to \citet{Phantom} for further details on the code.

We initialized the disk based on the grid-based simulation setup described in Section \ref{sec:initial_setup}. However, due to the intrinsic differences between grid-based and SPH codes, the exact setup differs inevitably and we focus on describing these differences below.
We used 10 million SPH particles, initially placed between $r=0.3$ and $r=3$. Compared to grid-based simulations, SPH simulations do not define a mesh, so particles can freely pass the initial inner/outer radii. We employed a fixed inner boundary at $r=0.2$ inside of which particles are removed from the simulation. This, and the spreading of the outer disk, means that the mass (and resolution) enclosed between $r=0.3$ and $r=3$ depletes with time, in contrast to the damping boundary conditions employed in the grid-based codes where arbitrary amounts of mass are continuously injected or removed in the damping zones to maintain the initial density profile in the damping zone. We found that about 30\% of the initial disk mass is lost through the inner boundary after 500 orbits. Unlike some previous \Phantom\ simulations where the mass lost from the disk is added to the star, we fixed the stellar mass to be consistent with grid-based simulations. Another difference compared with grid-based simulations is that the numerical resolution of SPH simulations depends on the local density and thus varies across the simulation domain and over time. This density dependency allows the best numerical resolution in the regions of high density, such as the disk midplane, while the decrease in the density, which can happen near the surface or within a planet-carved gap, results in poor numerical resolution.

We embedded a planet in the disk orbiting at $r=1$, having a mass $M_p = 10^{-3}M_*$. The potential of the planet was ramped up over the initial 100 orbits, consistently with grid-based simulations. The planet was implemented as a prescribed potential consistently with grid-based simulations; however, note that this is different from the usual approach in  \Phantom\  simulations where sink particles are used to model planets in a way that explicitly conserves momentum  \citep{bate1995}. Using a prescribed potential fixes the planet on a circular orbit and does not allow for orbital migration.

We added a fixed kinematic viscosity of $\nu = 10^{-5}$ in code units, following the `two first derivatives' formulation of the Navier Stokes terms outlined in Section 3.2.4 of \citet{lodato2010}. This is in addition to the usual shock capturing terms which are implemented with the \citet{cullen2010} switch with $\alpha_{\rm av} \in [0,1]$ and $\beta_{\rm av} = 2$ as outlined in \citet{Phantom}.

\subsection{Results}
\label{sec:hydro_results}

\begin{figure*}
    \centering
    \includegraphics[width=\textwidth]{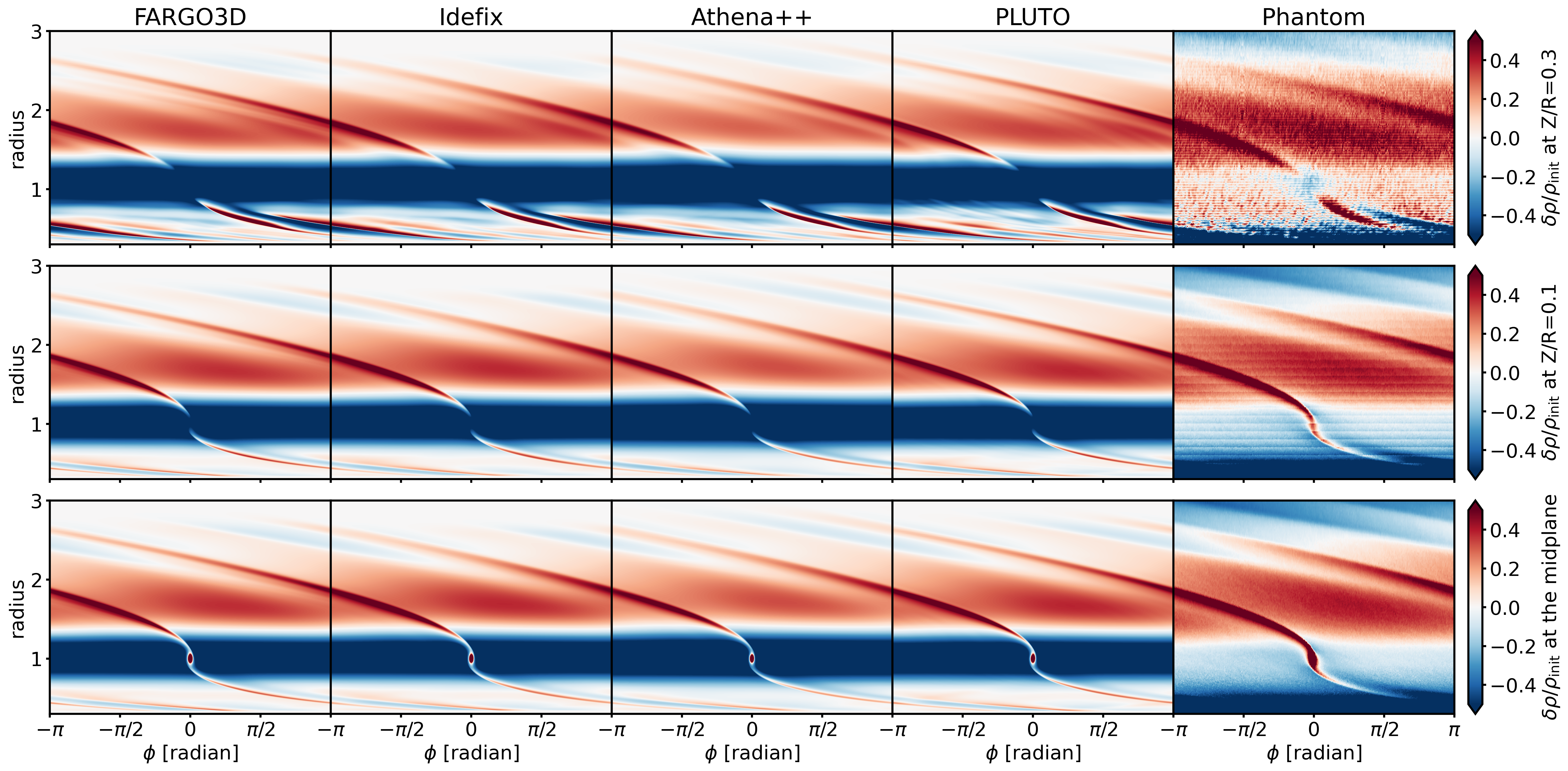}
    \caption{Perturbed density (normalized by the initial density), $\delta\rho/\rho_{\rm init}$ at (top panels) $Z/R=0.3$ (3 scale heights above the midplane at $R=1$), (middle panels) $Z/R=0.1$ (1 scale height above the midplane at $R=1$), and (bottom panels) midplane. From left to right, results with \texttt{FARGO3D}, \texttt{Idefix}, \texttt{Athena++}, \texttt{PLUTO}, and \texttt{Phantom}. At high altitudes, \Phantom\ results appear noisy because of the small number of SPH particles in the low-density regions.}
    \label{fig:dens2d}
\end{figure*}

\begin{figure*}
    \centering
    \includegraphics[width=\textwidth]{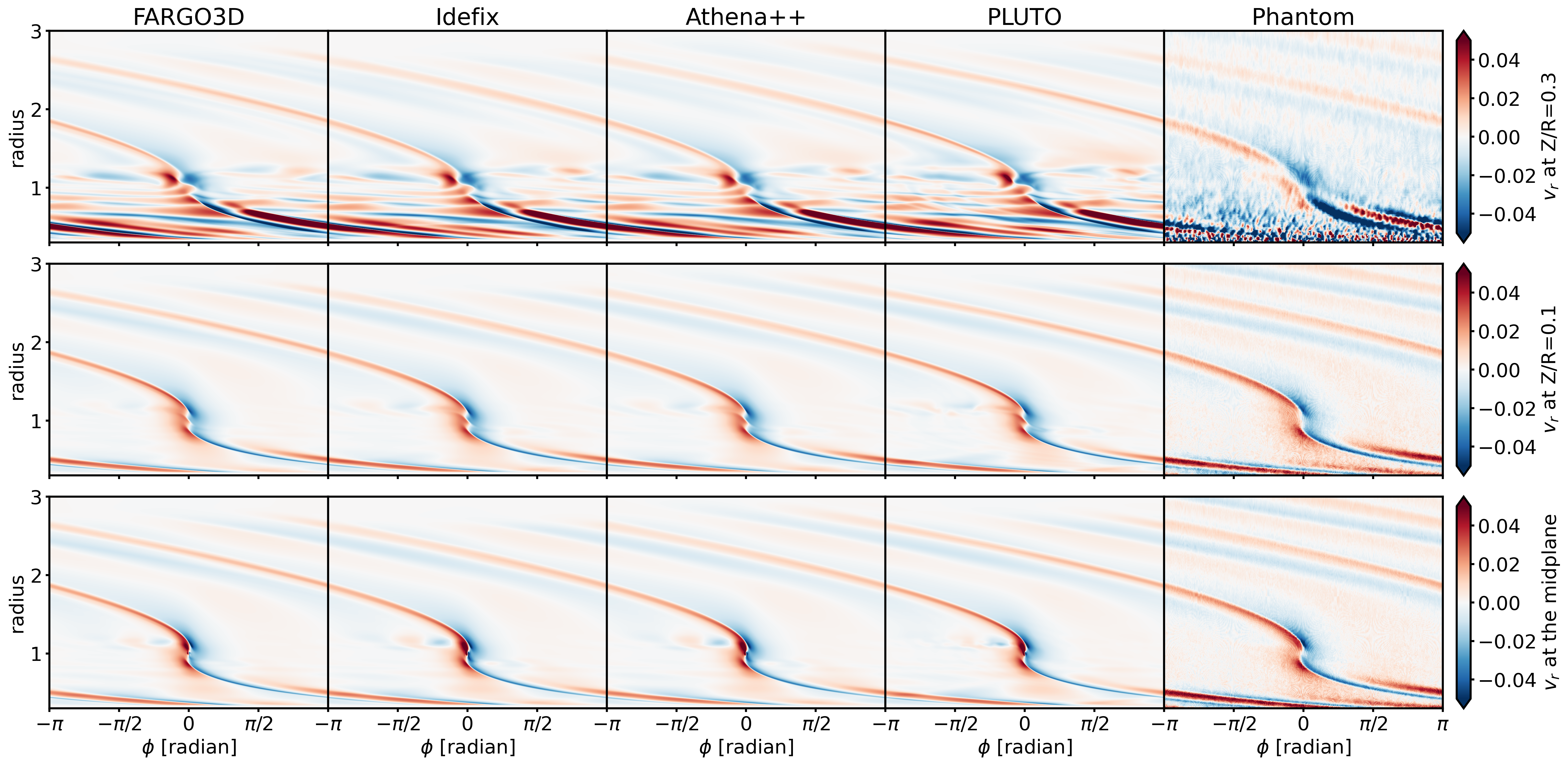}
    \caption{Same as Figure \ref{fig:dens2d}, but for the radial velocity $v_r$.}
    \label{fig:vrad2d}
\end{figure*}

\begin{figure*}
    \centering
    \includegraphics[width=\textwidth]{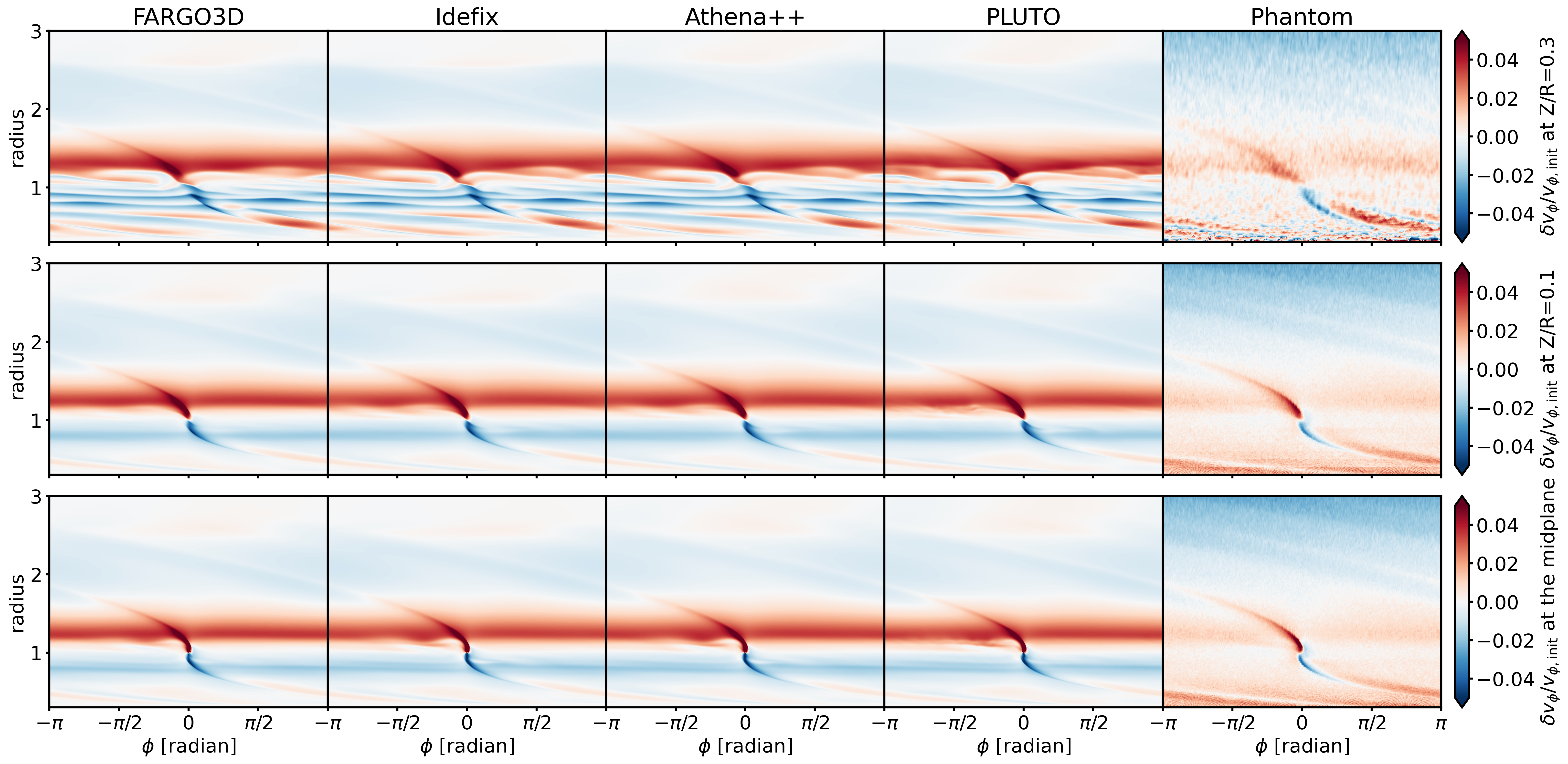}
    \caption{Same as Figure \ref{fig:dens2d}, but for the perturbed azimuthal velocity $\delta v_\phi \equiv v_\phi - v_{\phi, \rm init}$, normalized by the initial azimuthal velocity $v_{\phi,{\rm init}}$.}
    \label{fig:vphi2d}
\end{figure*}

\begin{figure*}
    \centering
    \includegraphics[width=\textwidth]{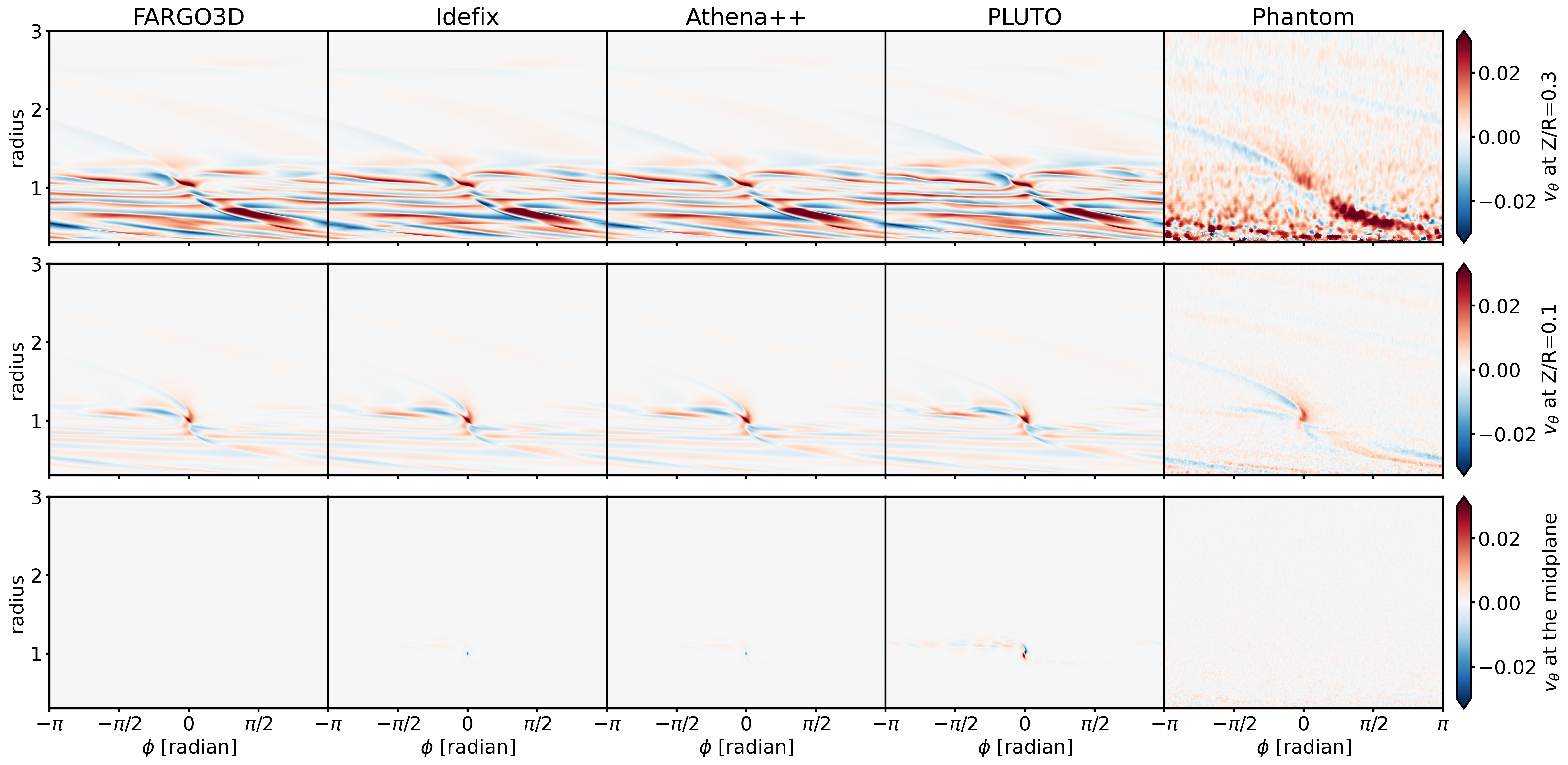}
    \caption{Same as Figure \ref{fig:dens2d}, but for the meridional velocity $v_\theta$.}
    \label{fig:vtheta2d}
\end{figure*}

In Figures \ref{fig:dens2d} -- \ref{fig:vtheta2d}, we present two-dimensional contour plots in a $\phi - r$ plane, showing the gas density, radial velocity, azimuthal velocity, and meridional velocity at three different heights after 500 planetary orbits of evolution: $Z/R=0, 0.1$, and 0.3. These heights are chosen to cover a range of vertical layers which various molecular lines having different optical depths probe (\citealt{Galloway_exoALMA}; see also review by \citealt{miotello2023} and references therein). For a more quantitative comparison, we present one-dimensional azimuthal distributions of the gas density, radial velocity, azimuthal velocity, and meridional velocity in Appendix \ref{sec:1dplots}.

\begin{figure*}
    \centering
    \includegraphics[width=\textwidth]{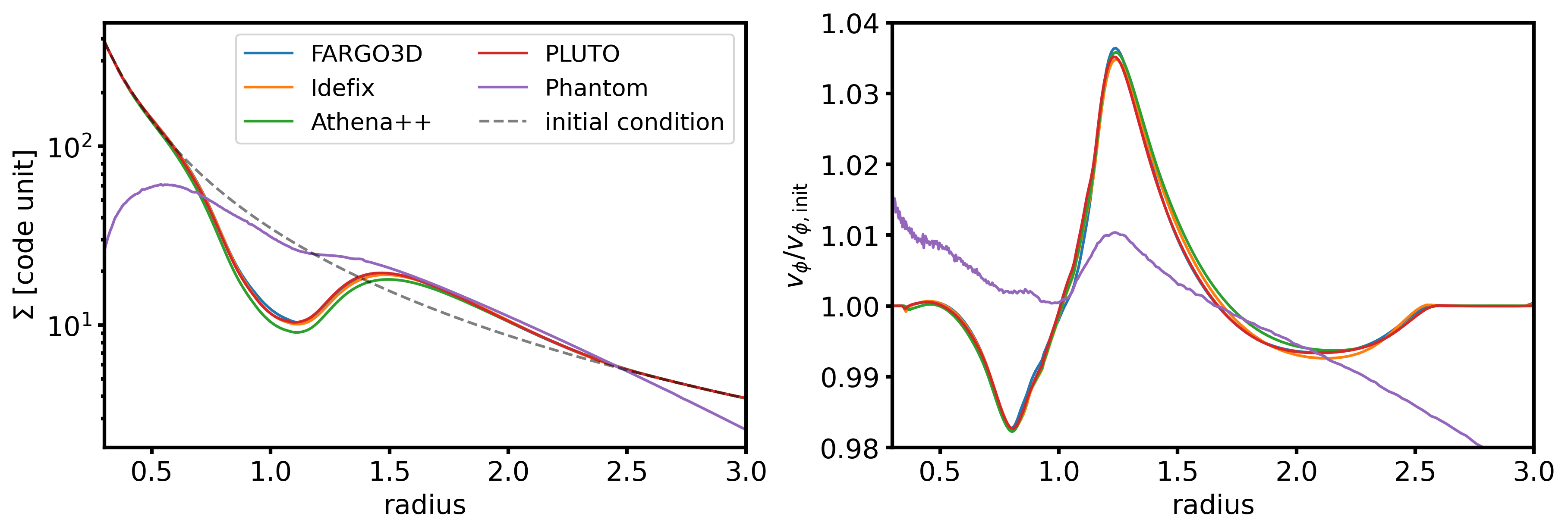}
    \caption{Radial profiles of the (left) surface density $\Sigma$ and (right) azimuthal velocity at the midplane $v_\phi$, normalized by the initial azimuthal velocity $v_{\phi, {\rm init}}$. While we present the azimuthal velocity at the midplane only, we confirm that the velocity at other heights show similar trend and good agreement between grid-based codes, with smaller velocity perturbations near the planet in the $\Phantom$ simulation due to the shallower gap.}
    \label{fig:radial_profiles}
\end{figure*}

Looking first at the perturbed density presented in Figure \ref{fig:dens2d}, all the grid-based simulations show an excellent agreement with each other although the level of density perturbation may differ by up to a few percent in the upper layers. Because particles can freely flow through the inner and outer boundary, the \Phantom\ simulation shows lower density near the radial boundaries. At the inner disk in particular, the low density results in a higher numerical viscosity. Together with strong shock dissipation in the inner disk, this leads to a rapid depletion which in turn flattens the surface density profile and makes the planet-opened gap shallower.  
We caution that inferring the mass of gap-opening planet based on the gap shape from \Phantom\ simulations could result in an overestimation. Nevertheless, we emphasize that the overall morphology of the spirals driven by the planet and the level of density perturbation associated with the spirals seen in the \Phantom\ simulation agree well with those seen in grid-based simulations, presumably because they are not sensitive to the underlying surface density profile \citep{ogilvie2002,rafikov2002,baezhu2018}.

Similarly to the density perturbation, velocity perturbations associated with the spirals show agreement among grid-based simulations (Figures \ref{fig:vrad2d} -- \ref{fig:vtheta2d}). In the \Phantom\ simulation, the perturbation in the azimuthal velocity across the gap is of the order of $1\%$ of the Keplerian speed, much smaller than that in grid-based simulations (Figure \ref{fig:radial_profiles}), due to the shallower gap. However, the velocity perturbations along the spirals in the \Phantom\ simulation are comparable to those in grid-based simulations. Because the strength of velocity kinks and Doppler flips is closely related to the level of velocity perturbations along the spirals, the consistency across all five hydrodynamics codes is important in the retrieval of the location of the planet, as we will show later in Section \ref{sec:discminer}.

For a more quantitative comparison of the gap structure between the models, in Figure \ref{fig:radial_profiles} we present azimuthally averaged radial profiles of the surface density and azimuthal velocity. In grid-based simulations, the surface density at the center of the gap opened by a Jupiter-mass planet is suppressed by $\simeq 70~\%$. On the other hand, in the \Phantom~simulation, the surface density is decreased by only $\simeq 10~\%$. The shallow gap results in an azimuthal velocity variation across the gap of $\sim1~\%$, much smaller than $\sim6~\%$ seen in grid-based codes.

\section{Radiative Transfer Calculations} 
\label{sec:RT}

Radiative transfer calculations were carried out using  \mcfost\ version 4.1.08\footnote{\url{https://github.com/cpinte/mcfost/releases/tag/v4.1.08}} \citep{Pinte2006,Pinte2009} and \radmc\ version 2.0\footnote{\url{https://github.com/dullemond/radmc3d-2.0} (git commit ID: d80a15e)}\citep{radmc}, using the outputs at 500~planetary orbits from the hydrodynamic simulations introduced in the previous Section.
As part of the exoALMA collaboration, the \mcfost\ code was updated to natively read the output files from all hydrodynamic codes. We tested all the combinations between hydrodynamic and radiative transfer codes, except for the \Phantom\ + \radmc\ combination because at the time we write this paper, there is no \radmc\ module which allows one to read in \Phantom\ outputs and map SPH particles onto a structured \radmc\ grid. This results in nine total combinations.

\begin{figure*}
    \centering
    \includegraphics[width=\textwidth]{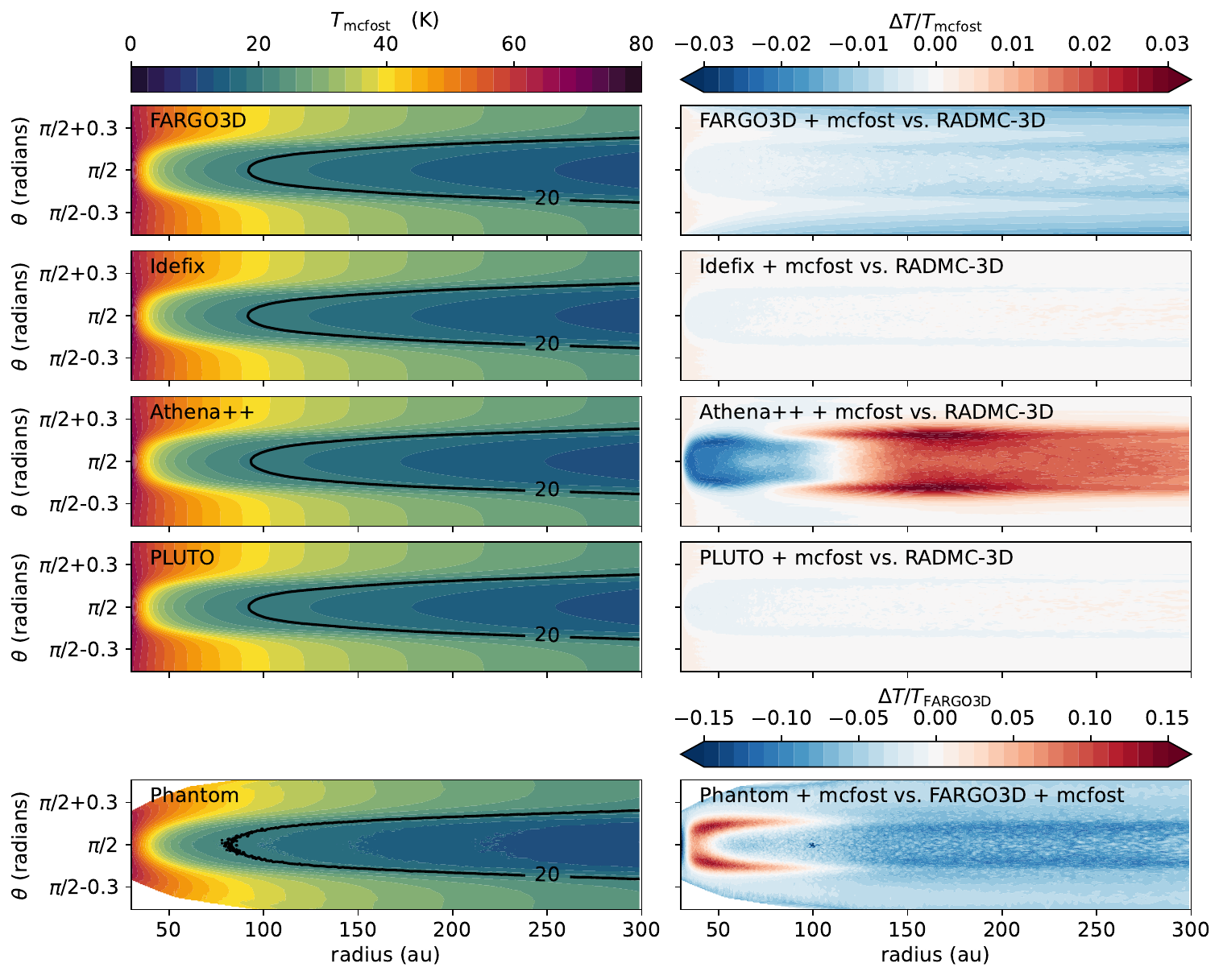}
    \caption{Azimuthally averaged temperature profiles in a $r-\theta$ plane from various combinations of hydrodynamic and radiative transfer codes (shown in the upper left corner of each panel). The black contours show where $T=20$~K. The white area in the upper left and lower left corner of the bottom panels is due to the lack of SPH particles there.}
    \label{fig:temp_comp}
\end{figure*}

\subsection{Radiative Transfer Calculation Setup}

We set the radiative transfer grid to be identical to that of grid-based hydrodynamic simulations described in Section \ref{sec:simulation_domain}. For \Phantom\ outputs, we used a Voronoi tessellation where each \mcfost\ cell corresponds to a SPH particle in order to avoid interpolating the density structure between the SPH and radiative transfer codes. 

To convert physical quantities from dimensionless hydrodynamic simulations into physical units for radiative transfer calculations, we placed the planet at 100~au from the solar-mass central star. The total initial gas mass in the disk was set to $0.01~M_\odot$. With this conversion, the initial gas surface density at the radial location of the planet is $0.52~{\rm g~cm}^{-2}$, which is broadly consistent with the gas surface density inferred for exoALMA disks \citep{Rosotti_exoALMA}. The star was placed at 140~parsecs from the observer.

To generate synthetic $^{12}$CO $J=3-2$ line cubes, we first determined the dust temperature by running Monte Carlo simulations. We placed a solar-mass star having a radius of $2~R_\odot$ and a temperature of 4000~K at the center of the domain, which emits a total of $10^9$ photons that interacts with dust grains. Dust opacities were calculated assuming a power-law grain size distribution following d$n(a)/{\rm d}a \propto a^{-3.5}$, between $a_{\rm min} = 0.03~\mu$m  and $a_{\rm max} = 1$~mm. The total dust mass was assumed to be $1\%$ of the gas mass in each grid cell. Grain composition was assumed to be astrosilicates \citep{Weingartner2001}. We computed the dust optical properties using the Mie theory, considering isotropic scattering only. The same opacities were used for both \mcfost\ and \radmc\ Monte Carlo simulations.

We assumed that the gas and dust are in local thermodynamic equilibrium, sharing the same temperature. We adopted a $^{12}$CO abundance of $10^{-4}$ relative to the total gas (by number) when the temperature is higher than 20~K. Below 20~K, we assumed that all CO molecules are frozen onto grains and no gas phase CO is present. We made an assumption that the local line width is set by thermal broadening only and ignored any non-thermal broadening. 

The disk inclination was set to 45 degrees, with a position angle of 90 degree (from north to east). The disk was placed at 140~pc and positioned such that the planet is located 45 degrees west of south.

\subsection{Results}
\label{sec:RT_results}

\subsubsection{Temperature}
\label{sec:rt_temperature}

The left panels of Figure \ref{fig:temp_comp} present azimuthally averaged temperature distributions of the five \mcfost\ models in a $r-\theta$ plane. The overall temperature structure exhibits a radially decreasing temperature profile, with a hotter atmosphere and a colder midplane, which is consistent with disk models where stellar irradiation is the dominating heating source \citep{chiang1997,dalessio1998}.

\begin{figure*}
    \centering
    \includegraphics[width=\textwidth]{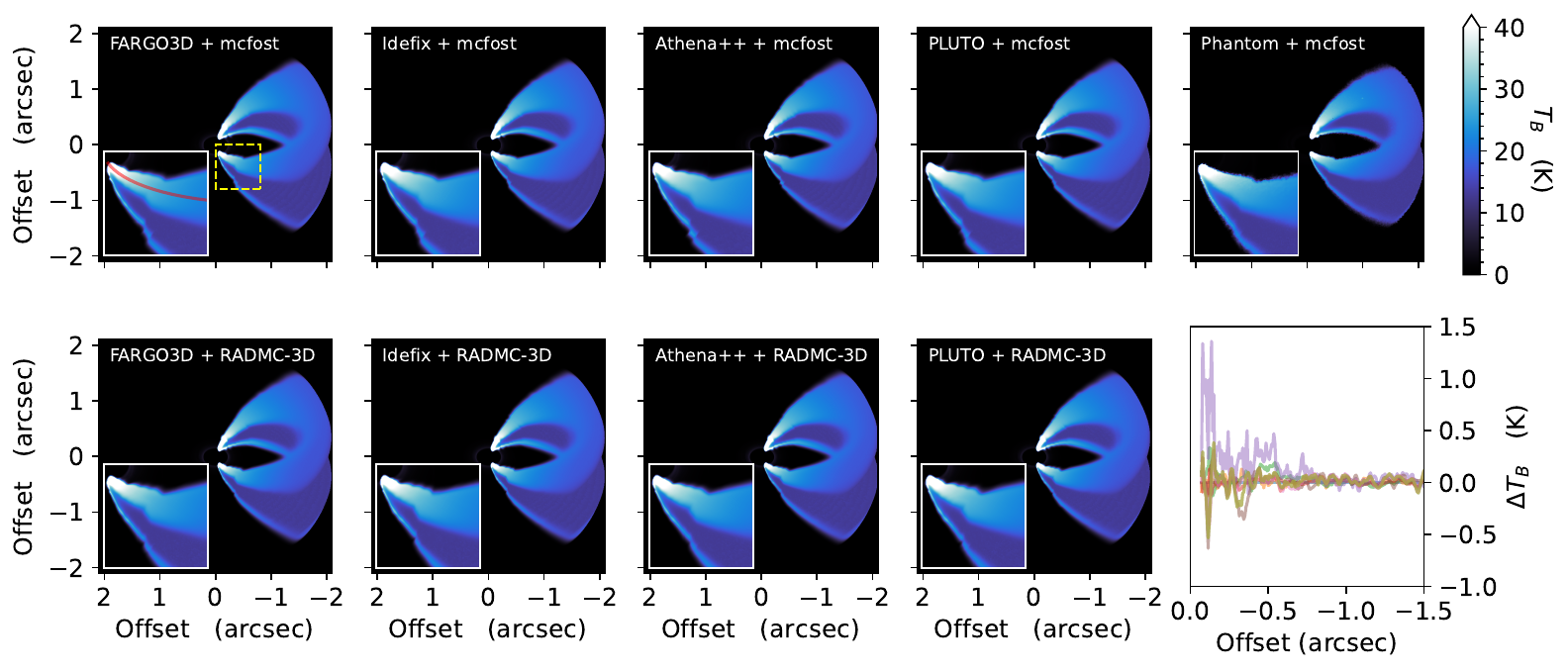}
    \caption{$^{12}$CO channel maps at $v_{\rm los} = 1.1~{\rm km~s}^{-1}$, using various combinations of hydrodynamic and radiative transfer codes (code names shown in the upper left corner of each panel). The planet, embedded and thus invisible in the channel maps, is located 45 degrees west of south. In all models, velocity kinks (indicated with a yellow dashed square in the upper left panel) are apparent in the vicinity of the planet. The inset in the lower-left quadrant of each panel shows a zoom-in view of the velocity kink. In the inset of the top left panel, we present the iso-velocity contour with a red curve. The bottom right panel shows the brightness temperature along the iso-velocity contour shown in the top left panel, compared to that of the \fargo\ + \mcfost\ model. Note that the brightness temperature of all models agree well with each other.} 
    \label{fig:rt_comp}
\end{figure*}

The temperature distributions show an excellent agreement within the four \mcfost\ models using outputs from the grid-based hydrodynmics codes. When the azimuthally averaged two-dimensional temperature distributions are compared grid by grid, no two models have a temperature difference greater than 0.84~K (between \idefix\ + \mcfost\ and \athena\ + \mcfost), with a domain-averaged mean difference ranging from 0.013~K (between \fargo\ + \mcfost\ and \pluto\ + \mcfost) to 0.18~K (between \idefix\ + \mcfost\ and \athena\ + \mcfost). In terms of fractional difference $\Delta T/T$, no two models have a difference greater than $2.7~\%$ (between \idefix\ + \mcfost\ and \athena\ + \mcfost) when compared grid by grid, with a domain-averaged mean difference ranging from $0.04~\%$ (between \fargo\ + \mcfost\ and \idefix\ + \mcfost) to $0.65~\%$ (between \idefix\ + \mcfost\ and \athena\ + \mcfost).

Similarly, an excellent agreement was found among the \radmc\ models. Based on grid-by-grid comparisons, we found no two models have a difference greater than 0.51~K (between \fargo\ + \radmc\ and \athena\ + \radmc). In terms of fractional difference, no two models have a difference greater than $1.8~\%$ (between \fargo\ + \radmc\ and \athena\ + \radmc).

When \mcfost~models are compared with \radmc~models, we found a strong consistency between the models. \idefix\ and \pluto\ models show an excellent agreement in particular; when the temperatures are compared grid by grid, these models differ by only $\lesssim 0.08~\%$ on average. The \athena\ models show the largest differences, but even in this case the fractional temperature difference is smaller than $3~\%$ everywhere in the domain. It is certainly worthwhile  further investigating why \athena\ models show larger differences but as we will show in the next section, the corresponding differences in $^{12}$CO channel maps are smaller than the typical noise in the exoALMA data and we defer such an investigation to a future study.

Lastly, we compared the \Phantom\ + \mcfost\ model to other \mcfost\ models. As an example, we present the difference between the temperature profiles from the \Phantom\ + \mcfost\ model and \fargo\ + \mcfost\ model in the bottom right panel of Figure \ref{fig:temp_comp}. As shown, the \Phantom\ model shows noticeable differences from other models, up to $\pm 15~\%$ in individual grid cells, with a domain-averaged mean difference of about $5~\%$. Given the differences in the density structure we discussed in Section \ref{sec:hydro_results}, this difference is not too surprising. Yet, this temperature difference results in only small differences in $^{12}$CO channel maps and may not be noticeable in exoALMA-quality data, as we will show now.

\subsubsection{Channel Maps}
\label{sec:rt_channel_map}

Figure \ref{fig:rt_comp} shows synthetic $^{12}$CO emission at a representative velocity channel. Overall, the brightness temperature at the surface shows visually negligible differences among the models. In addition, all synthetic cubes show clear and consistent velocity kinks in the vicinity of the planet, which is crucial for retrieving planet mass and location. In the bottom right panel of Figure \ref{fig:rt_comp}, we present the brightness temperature of each model relative to the reference model (\fargo\ + \mcfost), along the iso-velocity contour shown in the top left panel. As can be seen from the figure, all models agree with each other within $\pm 1.5$~K, with the largest difference seen from the \Phantom\ + \mcfost\ model (lavender color). At the location of the planet (offset $\simeq -0\farcs5$), the difference among the models is  $<0.5$~K. Given that the typical level of noise in the exoALMA molecular line images is about 1.5~K \citep{Teague_exoALMA}, we argue that the small, sub-Kelvin differences in brightness temperature among our models should not be a concern when it comes to modeling/finding embedded protoplanets.

\begin{figure*}
    \centering
    \includegraphics[width=1.0\textwidth]{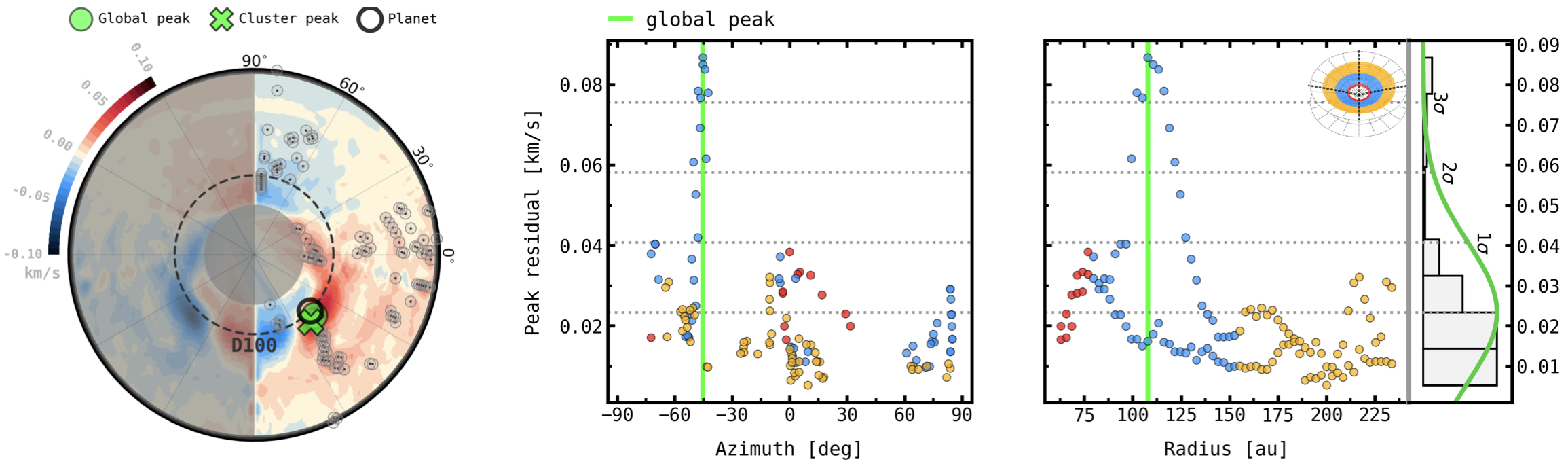}
    \caption{(Left panel) Folded velocity residual maps \citep[see][]{izquierdo2023,Izquierdo_exoALMA} computed for the $^{12}$CO synthetic cube created with \fargo\ and \mcfost, as an example. Green circle and cross indicate the location of peak and clustered residuals, whereas the black circle shows the correct planet location (100~au and $-45$ deg). (Middle panel) Peak residual velocity as a function of azimuthal angle. (Right panel) Peak residual velocity as a function of radius.}
    \label{fig:discminer_mcfost}
\end{figure*}

\begin{figure*}
    \centering
    \includegraphics[width=1.0\textwidth]{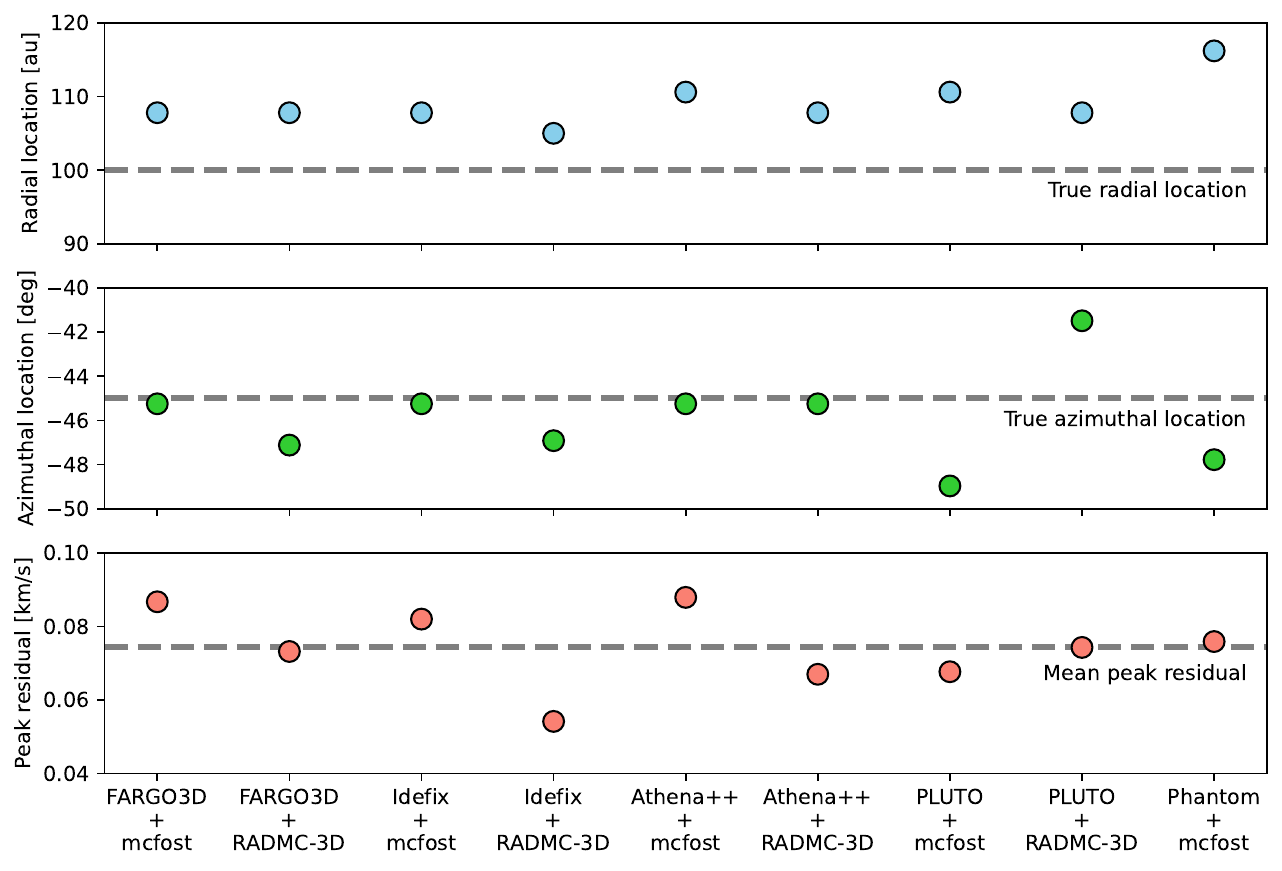}
    \caption{The (top) radial and (middle) azimuthal location of the planet inferred by \discminer\ (global peak in Figure \ref{fig:discminer_mcfost}) for synthetic $^{12}$CO cubes generated with different combinations of hydrodynamic and radiative transfer codes. The bottom panel shows the peak residual velocity at the inferred planet location.}
    \label{fig:discminer_comparison}
\end{figure*}

\section{Retrieval of Planet Location using \discminer}
\label{sec:discminer}

We used \discminer\footnote{\url{https://github.com/andizq/discminer}} to retrieve the planet location in synthetic cubes. To do so, we convolved the raw synthetic cubes with a $0\farcs15$ Gaussian beam and added random noises with a standard deviation of 1.5~K in each $100~{\rm m~s}^{-1}$ channel; these are the beamsize and level of noise of fiducial exoALMA molecular line images \citep{Teague_exoALMA}. For the detailed workflow of the retrieval, we refer the reader to \citet{discminer,izquierdo2023}. 

In Figure \ref{fig:discminer_mcfost}, we present a folded velocity residual map for the $^{12}$CO synthetic cube generated with \fargo\ and \mcfost, along with peak residual velocity profile as a function of azimuthal angle and radius. \discminer\ found a peak in the folded velocity residual map with a $3.64\sigma$ significance, and inferred the planet to be at 107.8~au in radius and $-45.25$~degrees in azimuth, which are in a good agreement with the true planet location: 100~au and $-45$~degrees.

To compare the inferred location of the planet and the associated peak velocity residual, we plot these quantities in Figure \ref{fig:discminer_comparison}.
Overall, the radial and azimuthal locations of the planet are  consistent across the models and agree well with the true planet location. However, we found that the inferred radial locations are systematically biased toward slightly large radial distances with a mean distance of 109.0~au, a minimum distance of  105.0~au (\idefix\ + \radmc), and a  maximum distance of 116.2~au (\Phantom\ + \mcfost). This bias in radial location was previously noted in \citet{discminer}, too, and we speculate that this bias is because the peak residual is generally more sensitive to the outer spiral arm. Note, however, that the mean retrieved radial location is within one beam ($0\farcs15$, or 21~au at 140~pc) from the true location. In case of the inferred azimuthal location, they are clustered around the true location with a mean azimuthal angle of $-45.92$~degrees, a minimum azimuthal angle of $-48.97$~degrees (\pluto\ + \mcfost), and a maximum azimuthal angle of $-41.49$~degrees (\pluto\ + \radmc).
Lastly, we found that the peak residual velocity is broadly consistent within about $20~\%$ variations among the models, with a mean of $0.074~{\rm km~s}^{-1}$, a minimum of $0.054~{\rm km~s}^{-1}$ (\idefix\ + \radmc), and a maximum of $0.088~{\rm km~s}^{-1}$ (\athena\ + \mcfost).

In summary, synthetic observations produced by any combination of the hydrodynamics and radiative transfer codes tested in this work exhibit broadly consistent localized perturbations near the true planet location.

\section{Conclusion}
\label{sec:conclusion}

To demonstrate the consistency of the codes used for forward modeling protoplanetary disks from the community, we conducted a systematic comparison of various hydrodynamics and radiative transfer codes. To do so, we first carried out three-dimensional planet-disk interaction simulations using four grid-based hydrodynamics codes (\fargo, \idefix, \athena, \pluto) and a smoothed particle hydrodynamics code (\Phantom). Overall, we found an excellent agreement between grid-based codes, in terms of the density and velocity structure. However, due to the different boundary conditions in the \Phantom\ simulation, we found that the planet-opened gap is shallower than those in grid-based simulations, which results in smaller azimuthal velocity deviations from the Keplerian rotation. Despite the difference in the gap depth, the overall morphology of the planet-driven spirals and the level of density/velocity perturbations associated with the spirals in the \Phantom\ simulation are in a good agreement with those in other hydrodynamics codes. 

We used two radiative transfer codes (\mcfost, \radmc) to calculate the disk temperature and generate synthetic $^{12}$CO cubes. The disk temperatures showed a good agreement within \mcfost\ cubes and \radmc\ cubes separately, as well as between \mcfost\ and \radmc\ models based on the same hydrodynamic simulation. Due to the different density structure, the disk temperature in the \Phantom\ + \mcfost\ model is different from other models up to $\pm15~\%$ in individual grid cells, with a domain-averaged mean difference of about $5~\%$. However, the corresponding difference in $^{12}$CO emission is $\leq 1.5$~K, within the typical noise in the exoALMA data.

Using \discminer, we retrieved the location of the planet from the synthetic cubes. Despite some differences in the hydrodynamics and/or radiative transfer calculation results, the retrieved planet locations are consistent across all cubes and agree well with the true planet location. This is because the density and velocity perturbations along the planet-driven spirals, which determines the location and strength of the velocity kink in channel maps, are consistent across all the codes we tested. 

In summary, our findings suggest that any combination of the hydrodynamics and radiative transfer codes we tested in this paper could be used to model and interpret planet-driven kinematic perturbations.
Admittedly, we only considered a single base hydrodynamics model. Future studies are needed to confirm that the consistency across all codes holds when changes are made to the model, such as planet mass, orbital eccentricity/migration, vertical temperature stratification, finite gas cooling time, disk viscosity, and the presence/strength of disk's magnetic fields.

\section{Acknowledgments}

This Letter makes use of the following ALMA data: ADS/JAO.ALMA\#2021.1.01123.L. ALMA is a partnership of ESO (representing its member states), NSF (USA) and NINS (Japan), together with NRC (Canada), MOST and ASIAA (Taiwan), and KASI (Republic of Korea), in cooperation with the Republic of Chile. The Joint ALMA Observatory is operated by ESO, AUI/NRAO and NAOJ. The National Radio Astronomy Observatory is a facility of the National Science Foundation operated under cooperative agreement by Associated Universities, Inc. We thank the North American ALMA Science Center (NAASC) for their generous support including providing computing facilities and financial support for student attendance at workshops and publications. 
Part of the calculations were performed on the OzSTAR national facility at Swinburne University of Technology. The OzSTAR program receives funding in part from the Astronomy National Collaborative Research Infrastructure Strategy (NCRIS) allocation provided by the Australian Government, and from the Victorian Higher Education State Investment Fund (VHESIF) provided by the Victorian Government.
JB acknowledges support from NASA XRP grant No. 80NSSC23K1312. MB, DF, JS have received funding from the European Research Council (ERC) under the European Union’s Horizon 2020 research and innovation programme (PROTOPLANETS, grant agreement No. 101002188). Computations by JS have been performed on the `Mesocentre SIGAMM' machine, hosted by Observatoire de la Cote d’Azur. PC acknowledges support by the Italian Ministero dell'Istruzione, Universit\`a e Ricerca through the grant Progetti Premiali 2012 – iALMA (CUP C52I13000140001) and by the ANID BASAL project FB210003. NC has received funding from the European Research Council (ERC) under the European Union Horizon Europe research and innovation program (grant agreement No. 101042275, project Stellar-MADE). SF is funded by the European Union (ERC, UNVEIL, 101076613), and acknowledges financial contribution from PRIN-MUR 2022YP5ACE. MF is supported by a Grant-in-Aid from the Japan Society for the Promotion of Science (KAKENHI: No. JP22H01274). CH acknowledges support from NSF AAG grant No. 2407679. IH is supported by an Australian Government Research Training Program (RTP) Scholarship. Support for AFI was provided by NASA through the NASA Hubble Fellowship grant No. HST-HF2-51532.001-A awarded by the Space Telescope Science Institute, which is operated by the Association of Universities for Research in Astronomy, Inc., for NASA, under contract NAS5-26555. GL has received funding from the European Union's Horizon 2020 research and innovation program under the Marie Sklodowska-Curie grant agreement No. 823823 (DUSTBUSTERS). CL has received funding from the European Union's Horizon 2020 research and innovation program under the Marie Sklodowska-Curie grant agreement No. 823823 (DUSTBUSTERS) and by the UK Science and Technology research Council (STFC) via the consolidated grant ST/W000997/1. CP acknowledges Australian Research Council funding via FT170100040, DP18010423, DP220103767, and DP240103290. DP acknowledges Australian Research Council funding via DP18010423, DP220103767, and DP240103290. GR acknowledges funding from the Fondazione Cariplo, grant no. 2022-1217, and the European Research Council (ERC) under the European Union’s Horizon Europe Research \& Innovation Programme under grant agreement no. 101039651 (DiscEvol). GWF acknowledges support from the European Research Council (ERC) under the European Union Horizon 2020 research and innovation program (Grant agreement no. 815559 (MHDiscs)). GWF was granted access to the HPC resources of IDRIS under the allocation A0120402231 made by GENCI. TCY acknowledges support by Grant-in-Aid for JSPS Fellows JP23KJ1008. Views and opinions expressed by ERC-funded scientists are however those of the author(s) only and do not necessarily reflect those of the European Union or the European Research Council. Neither the European Union nor the granting authority can be held responsible for them.


\software{\fargo~\citep{fargo3d}, \idefix~\citep{idefix}, \athena~\citep{athena}, \pluto~\citep{pluto}, \Phantom~\citep{Phantom}, \mcfost~\citep{mcfost}, \radmc~\citep{radmc}, \discminer~\citep{discminer}}



\appendix

\section{Azimuthal Density and Velocity Profiles}
\label{sec:1dplots}

\begin{figure*}
    \centering
    \includegraphics[width=\textwidth]{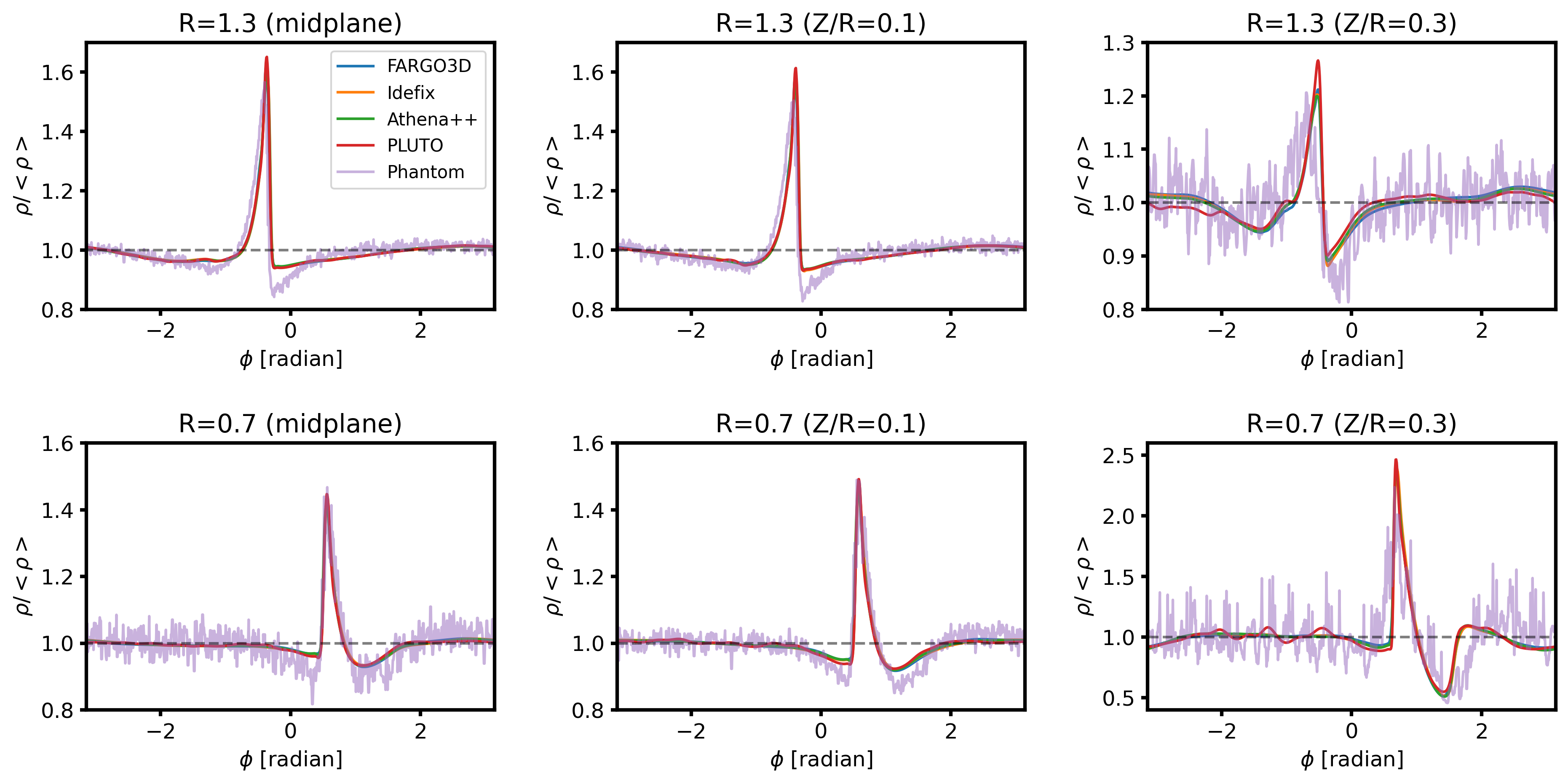}
    \caption{Azimuthal distribution of the density at (top panels) $R=1.3$ and (bottom panels) $R=0.7$. From left to right, each panel shows the density at the midplane, at $Z/R=0.1$, and $Z/R=0.3$, respectively. The scatter seen in the \Phantom~simulation is due to the finite number of particles used in the simulation, and is expected.}
    \label{fig:dens1d}
\end{figure*}

\begin{figure*}
    \centering
    \includegraphics[width=\textwidth]{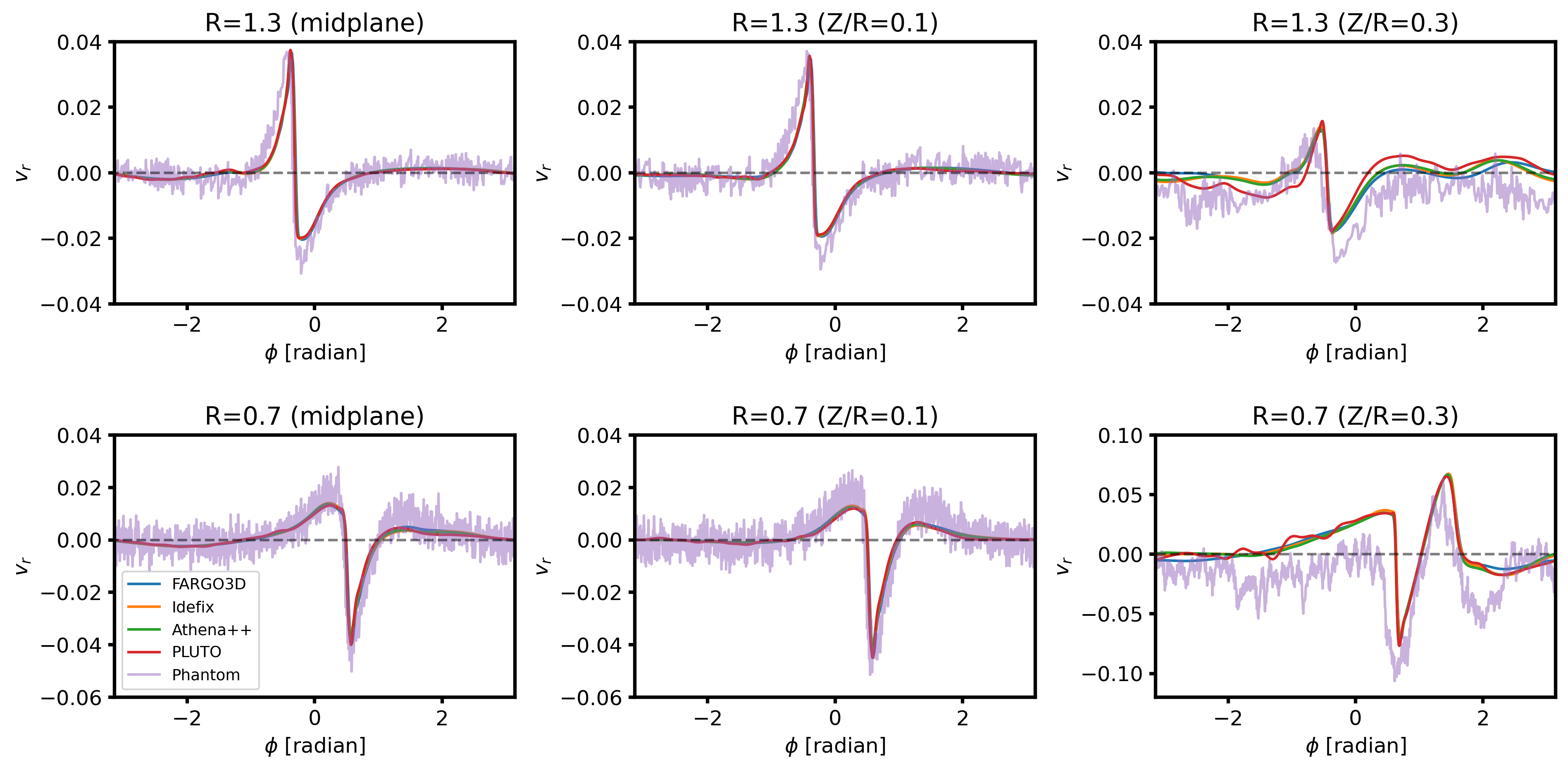}
    \caption{Same as Figure \ref{fig:dens1d}, but for the radial velocity $v_r$.}
    \label{fig:vrad1d}
\end{figure*}

\begin{figure*}
    \centering
    \includegraphics[width=\textwidth]{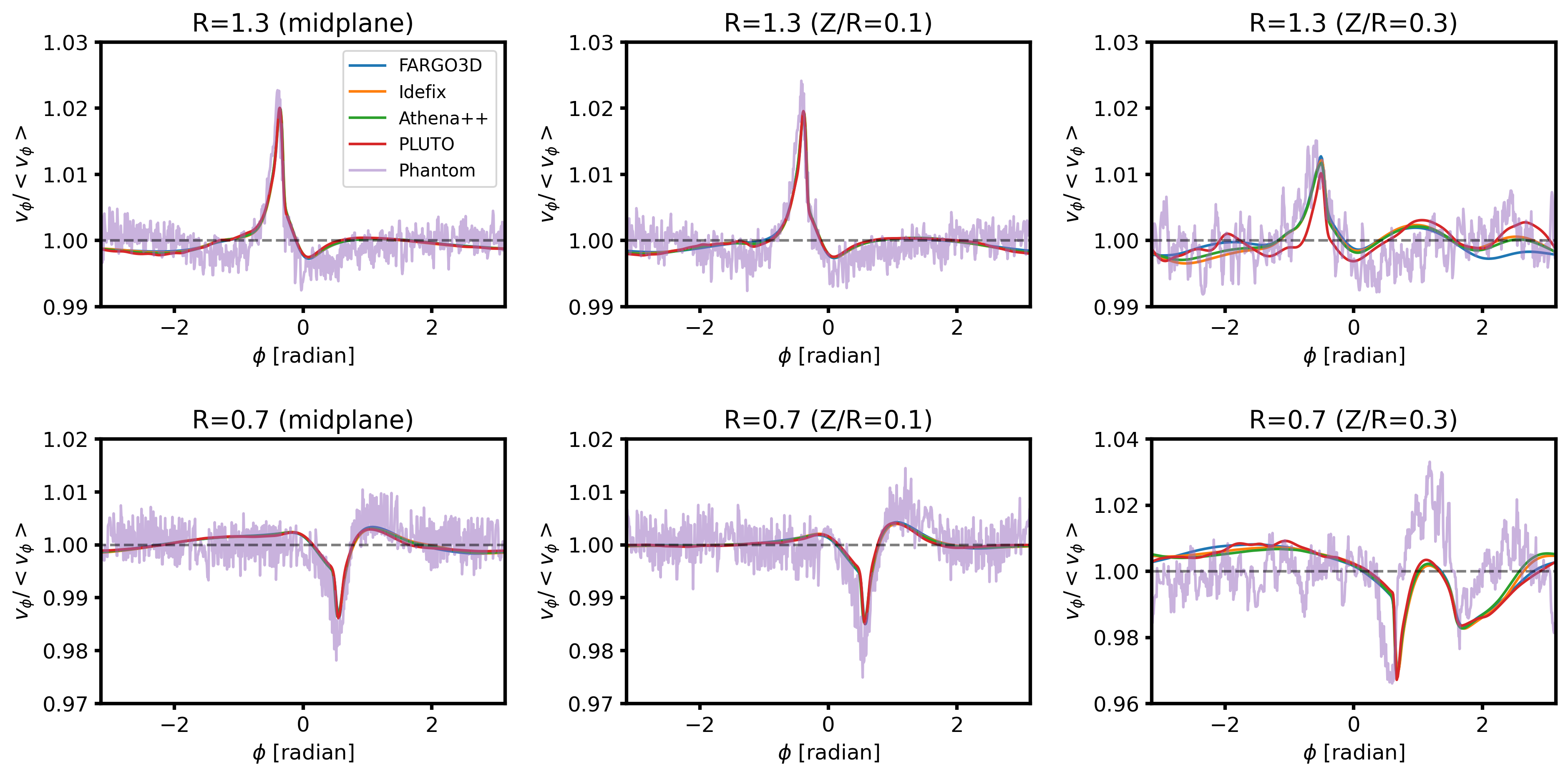}
    \caption{Same as Figure \ref{fig:dens1d}, but for the azimuthal velocity $v_\phi/\langle v_{\phi} \rangle$.}
    \label{fig:vphi1d}
\end{figure*}

\begin{figure*}
    \centering
    \includegraphics[width=\textwidth]{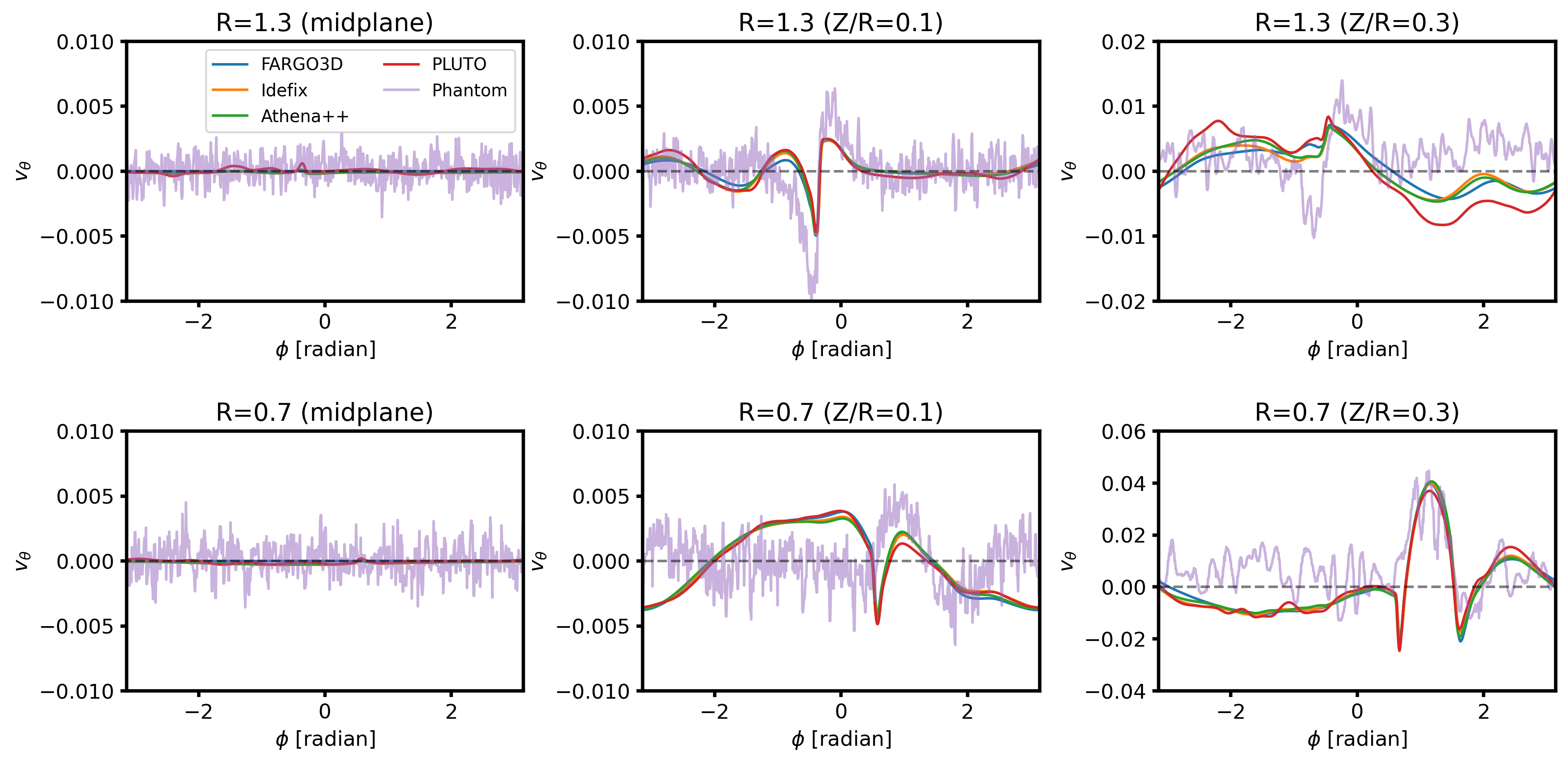}
    \caption{Same as Figure \ref{fig:dens1d}, but for the meridional velocity $v_\theta$.}
    \label{fig:vtheta1d}
\end{figure*}

In Figures \ref{fig:dens1d} -- \ref{fig:vtheta1d}, we present azimuthal distributions of the density, radial velocity, azimuthal velocity, and meridional velocity at $R=0.7$ and $R=  1.3$. In each figure, we present the azimuthal distributions at three different heights, $Z/R=0, 0.1$, and 0.3. 

\bibliography{bibliography}{}
\bibliographystyle{aasjournal}



\end{document}